\documentclass{ptephy_v1}
\usepackage{graphicx,color}
\usepackage{bm}
\usepackage{braket}
\usepackage{mathtools}
\usepackage{amsfonts}
\usepackage{ulem}
\usepackage{appendix}
\usepackage{amsthm}

\begin{document}
\title{Construction of interacting flat-band models by molecular-orbital representation: Correlation functions, energy gap, and entanglement}
\author{Tomonari Mizoguchi$^\ast$, Yoshihito Kuno, and Yasuhiro Hatsugai}
\affil{Department of Physics, University of Tsukuba, Tsukuba, Ibaraki 305-8571, Japan \email{mizoguchi@rhodia.ph.tsukuba.ac.jp}}

\begin{abstract}
We calculate correlation functions of exactly-solvable one-dimensional flat-band models
by utilizing the ``molecular-orbital" representation.
The models considered in this paper have a gapped ground state with flat-band being fully occupied, even in the presence of the interaction.
In this class of models, the space spanned by the ``molecular-orbitals" is the co-space of that spanned by the flat bands.
Thanks to this property, the correlation functions are calculated by using
the information of the molecular-orbitals rather than the explicit forms of the flat-band wave functions, 
which simplifies the calculations.
As a demonstration, several one-dimensional models and their correlation functions are presented. 
We also calculate the entanglement entropy by using the correlation function.
\end{abstract}
\subjectindex{146}

\maketitle
\section{Introduction}
Flat-bands of single-particle spectrum in lattice models have attracted considerable 
interests in condensed matter physics.
Vanishing of band width makes correlation effects dominant. 
In this line, various correlation-induced phenomena are investigated, such as ferromagnetism~\cite{Mielke1991,Mielke1991_2,Tasaki1992,Mielke1993,Tasaki1998,Tanaka2007,Tamura2019,Tanaka2020,Tasaki2020}, superconductivity~\cite{Imada2000,Miyahara2007,Tovmasyan2013,Peotta2015,Kobayashi2016,Aoki2020}, Bose-Einstein condensation~\cite{Huber2010,Katsura2021}, 
correlated topological phases~\cite{Katsura2010,Green2010,Tang2011,Sun2011,Neupert2011,Sheng2011,Regnault2011}, and so on. 
Recently, flat-band systems have also drawn interests in terms of dynamics~\cite{Vidal1998,Mukherjee2018,Kuno2020_Creutz} and localization~\cite{Goda2006,Chalker2010,Leykam2017,Kuno2020,Hatsugai2021}.
 
In flat-band models, complete quench of kinetic energy may allow us to obtain the exact ground state
by simply finding a state that minimizes the interaction term. 
The ferromagnetic state in the flat-band Hubbard model is a representative example.
Recently, this protocol was also applied to the Wigner crystal~\cite{Wu2007} and the quantum scar~\cite{Hart2020,Kuno2020_scar,Kuno2021}. 
In these examples, the interaction has a natural form (e.g., the Hubbard interaction or the nearest-neighbor density-density interaction),
but the flat-band wave functions have a suitable real-space profile 
so that the interaction terms do not act on the many-body states composed of the flat-bands. 

On the other hand, it is also possible to tune the interaction 
such that it is vanishing when acting on the many-body states with the flat-band states being occupied.
In this paper, we argue this type of interacting flat-band models.
For the construction of the models, we rely on the molecular-orbital (MO) representation,
which we have developed to describe flat-band models~\cite{Hatsugai2021,Hatsugai2011,Hatsugai2015,Mizoguchi2019,Mizoguchi2020,Mizoguchi2021}.
The key idea of the MO representation is to write down the single-particle Hamiltonian by using 
the unnormalized and non-orthogonal wave functions, i.e., the ``MOs", whose number is smaller than the number of atomic sites.
Based on this, we consider the interaction term written down by the MOs. 
The form of the interaction in the atomic-site basis is, in general, not in a natural form as the conventional interactions, but is highly fine-tuned. 
Nevertheless, the models possess several intriguing features. 
Namely, we not only obtain the exact ground state, but also calculate the correlation function for the ground state.
Specifically the correlation function can be obtained without explicitly deriving the flat-band wave functions.
Rather, the forms of the MO have all the information needed 
to calculate the correlation function, which is a unique feature of this type of models and has not yet been pointed out to our knowledge. 

The rest of this paper is structured as follows.
In Sec.~\ref{sec:MOrep}, we review the MO representation of the flat-band models.
In Sec.~\ref{sec:single}, we describe the details of the single-particle eigenvalues and eigenfunctions of the models. 
In Sec.~\ref{sec:int}, we introduce the interaction term to construct the exactly-solvable flat-band models. The form of the ground state is also shown. 
In Sec.~\ref{sec:corr}, we present the formal expression of the correlation function for the ground state.
In Sec.~\ref{sec:ex}, we discuss two concrete examples, namely, 
the saw-tooth lattice model and the diamond chain,
and show the explicit forms of the correlation functions. 
For the saw-tooth lattice model, we also show the entanglement entropy computed 
by using the correlation function.
Section~\ref{sec:summary} is devoted to a brief summary of this paper. 

\section{Molecular-orbital representation of flat-band models \label{sec:MOrep}}
We first review the MO representation of generic flat-band models~\cite{Hatsugai2021,Hatsugai2011,Hatsugai2015,Mizoguchi2019,Mizoguchi2020,Mizoguchi2021}.
This model construction method can be viewed as an extension of the ``cell construction",
which has been developed in the context of the ferromagnetism in the Hubbard-type models~\cite{Tasaki1992,Tasaki1998,Tanaka2020,Tasaki2020}.
Let us consider a spinless-fermion lattice model 
of $N$ sites:
\begin{align}
\hat{{\cal H}}_0 = \sum_{i,j=1}^M \hat{C}_i ^\dagger h_{ij} \hat{C}_j, \label{eq:ham_single}
\end{align}
where $\hat{C}_i$ is an annihilation operator of a MO $i$, ($i=1,\cdots,M$) as
\begin{align}
 \hat{C}_i = \bm{\psi}_i ^\dagger \hat{\bm{c}},  
\end{align}
where
\begin{align}
\hat{\bm{c}} = \left(
\begin{array}{c}
\hat{c}_1\\
\vdots \\
\hat{c}_N\\
\end{array}
\right), \label{eq:fermion_ao}
\end{align}
and
\begin{align}
\bm{\psi}_i = 
\left(
\begin{array}{c}
\psi_{i,1} \\
\vdots \\
\psi_{i,N} \\
\end{array}
\right),
\end{align}
are, respectively, the original spinless fermion operators 
and the $N$ component column vector of the coefficients,
and $h$ is a Hermitian $M\times M$ matrix.
To be more specific, $\hat{c}_i$ in Eq.~(\ref{eq:fermion_ao}) denotes the annihilation operator of the spinless fermion at site $i$,
satisfying $\{\hat{c}_i, \hat{c}_j^\dagger\} = \delta_{i,j}$ and $\{\hat{c}_i, \hat{c}_j \} =\{\hat{c}^\dagger_i, \hat{c}_j^\dagger\} =0$.
For later use, we define
\begin{align} 
\Psi := (\bm{\psi}_1,\cdots,\bm{\psi}_M),
\end{align}
which is an $N\times M$ matrix.
Note that $\hat{C}_i$ satisfies the anti-commutation relation:
\begin{eqnarray}
\left\{\hat{C}_i, \hat{C}_j^\dagger \right\} =  (\bm{\psi}_i ^\dagger)_n (\bm{\psi}_j)_m \left\{\hat{c}_n , \hat{c}_m^\dagger \right\} = [\mathcal{O}]_{i,j},\label{eq:anticommun_MO}
\end{eqnarray}
where 
$\mathcal{O}:=\Psi ^\dagger \Psi$ is the $M \times M$ matrix and is referred to as an overlap matrix.

The single-particle Hamiltonian of Eq.~(\ref{eq:ham_single}) can be written by using $\bm{c}$ as
\begin{align}
\hat{{\cal H}}_0  &= \hat{\bm{c}}  ^\dagger H_0 \hat{\bm{c}},
\end{align}
where
\begin{align}
  H_0 &= \sum_{i,j=1}^M \bm{\psi}_i h_{ij} \bm{\psi}_ j ^\dagger 
  = \Psi h \Psi ^\dagger.  \label{eq:ham_morep}
\end{align}
Using the following formula 
for $N\times M$ and $M\times N$ matrices $A_{NM}$ and $B_{MN}$ 
\begin{align*}
  \det_N(I_N +A_{NM}B_{MN})
  &=   \det_M(I_M +B_{MN}A_{NM}),  
\end{align*}
we obtain
\begin{align*}
  \det _N(\lambda I_N-H_0)
  &=   \det _N(\lambda I_N-\Psi h\Psi ^\dagger)
  \\
  &= \lambda ^N\det_N(I_N- \lambda  ^{-1} \Psi h\Psi ^\dagger)
  \\
  &= \lambda ^N\det_M(I_M- \lambda  ^{-1}  h\Psi ^\dagger \Psi)  
  \\
  &= \lambda ^{N-M}\det_M(\lambda I_M- h\Psi ^\dagger \Psi).
\end{align*}
Then if $N>M$, there are at least $N-M (>0)$ zero modes~\footnote{The lower bound of the number of the zero modes
is straightforwardly obtained from Eq.~(\ref{eq:ham_morep}), 
because the dimension of the kernel of $\Psi^\dagger$
as a linear map is equal to or greater than $N-M$. 
It can also be found from Eq.~(\ref{eq:ham_morep}) that the number of the zero modes is equal to the dimension of the kernel of $h \Psi^\dagger$.
}.
When applying this argument to the momentum space representation, 
one obtains flat-bands.
One may apply it to the random case in a real space as well~\cite{Hatsugai2021,Mizoguchi2019}.
We also note that additional zero modes appear when
\begin{align}
  \det_M h\Psi ^\dagger \Psi &= 0. \label{eq:add_zero}
\end{align} 
Equation (\ref{eq:add_zero}) is satisfied when 
$\det_M h = 0$, or 
$\det_M \mathcal{O} = 0$~\cite{Mizoguchi2019}.

\section{Single-particle eigenvalues and eigenstates \label{sec:single}}
In what follows, we assume that det$\mathcal{O} \neq 0$.

Using the MOs, we can also derive the eigenstates 
other than the zero modes.
Let $\Psi^\prime$ be the eigenmultiplet:
\begin{eqnarray}
H_0 \Psi^{\prime}  =& \Psi^{\prime} \mathcal{E}, \nonumber \\
\Psi^\prime  =& \Psi \Phi, \label{eq:def_psip}
\end{eqnarray}
with $\mathcal{E} = \mathrm{diag} \left( \varepsilon_1, \cdots,\varepsilon_M \right)$ and $\Phi$ being the $M \times M$ matrix. Note that $\Phi$ is not unitary.
For later use, we assume $ \varepsilon_1 \leq  \varepsilon_2 \cdots  \leq \varepsilon_M$.
From Eqs.~(\ref{eq:ham_morep}) and (\ref{eq:def_psip}), we have 
\begin{eqnarray}
\Psi h \mathcal{O} \Phi = \Psi \Phi \mathcal{E}.
\end{eqnarray}
By further operating $\Psi^\dagger$ from the left, 
we obtain
\begin{eqnarray}
\mathcal{O} h \mathcal{O} \Phi =  \mathcal{O} \Phi \mathcal{E}. \label{eq:mo_eigen}
\end{eqnarray}
Defining $h_{\psi}:= \mathcal{O}^{1/2} h \mathcal{O}^{1/2}$, Eq.~(\ref{eq:mo_eigen}) 
can be further deformed as
\begin{eqnarray}
h_{\psi} \Phi^{\prime} = \Phi^{\prime} \mathcal{E}, \label{eq:hpsi_eigen}
\end{eqnarray}
where $\Phi^{\prime}:= \mathcal{O}^{1/2} \Phi$.
Importantly, we can choose $\Phi^{\prime}$ as a unitary matrix
since $h_{\psi}$ is an Hermitian matrix.
This leads to 
\begin{eqnarray}
\Phi^{\prime \dagger} \Phi^{\prime} = \Phi^{\dagger} \mathcal{O} \Phi = I_M, \label{eq:orthonomal}
\end{eqnarray}
and thus 
\begin{eqnarray}
\Phi^{\dagger}  = \left(\mathcal{O} \Phi \right)^{-1}. \label{eq:phi_inv}
\end{eqnarray}

From  Eqs.~(\ref{eq:hpsi_eigen}) (\ref{eq:orthonomal}), and (\ref{eq:phi_inv}), 
we have
\begin{eqnarray}
h = \mathcal{O}^{-1/2} h_{\psi} \mathcal{O}^{-1/2} =  \mathcal{O}^{-1/2} \left(\Phi^{\prime} \mathcal{E} \Phi^{\prime \dagger} \right) \mathcal{O}^{-1/2}
= \Phi  \mathcal{E} \Phi^{\dagger}. 
\label{eq:h}
\end{eqnarray}
Substituting Eq.~(\ref{eq:h}) into Eq.~(\ref{eq:ham_morep}), we can write down the single-particle Hamiltonian as
\begin{eqnarray}
\hat{\mathcal{H}}_0 = \hat{\bm{c}}^{\dagger} \Psi \Phi  \mathcal{E} \Phi^{\dagger} \Psi^{\dagger} \hat{\bm{c}}
 = \hat{\bm{C}}^{\prime \dagger}\mathcal{E} \hat{\bm{C}}^{\prime},
\end{eqnarray}
where $\hat{\bm{C}}^{\prime}:= \Psi^{\prime \dagger} \hat{\bm{c}} = \Phi^{\dagger} \Psi^{\dagger}\hat{\bm{c}} $ 
is a set of orthogonal MOs. 
We note that 
\begin{eqnarray}
\hat{\bm{C}} = \mathcal{O} \Phi \hat{\bm{C}}^\prime, \label{eq:C_Cp}
\end{eqnarray}
holds, and that 
$\hat{\bm{C}}^{\prime}$ satisfies the anti-commutation relation:
\begin{eqnarray}
\left\{ \hat{C}_i^{\prime}, \hat{C}_{j}^{\prime \dagger} \right\} 
=\sum_{k,l,k^\prime,l} [\Phi^{\dagger}]_{i,k} [\Psi^{\dagger} ]_{k,l} [\Psi ]_{l^{\prime},k\prime} [\Phi]_{k\prime,j} \left\{ \hat{c}_l ,\hat{c}^{\dagger}_{l^\prime}  \right\}
= [\Phi^{\dagger} \mathcal{O} \Phi ]_{i,j} 
= \delta_{i,j}. \label{eq:cp_anti}
\end{eqnarray}
To obtain (\ref{eq:C_Cp}) and (\ref{eq:cp_anti}), we have used (\ref{eq:orthonomal}). 

Let us here note that $\Psi^{\prime}$ is orthonormalized as
\begin{eqnarray}
\Psi^{\prime \dagger}\Psi^{\prime} = \Phi^{\dagger} \Psi^{\dagger} \Psi \Phi = \Phi^{\dagger} \mathcal{O} \Phi 
= I_M,
\end{eqnarray} 
where we have used (\ref{eq:phi_inv}).
However, it is not complete generically, since
\begin{eqnarray}
P_{\rm MO} = \Psi^{\prime}\Psi^{\prime \dagger} =  \Psi \Phi \Phi^{\dagger} \Psi^{\dagger} =\Psi  \Phi (\mathcal{O}\Phi)^{-1} \Psi^{\dagger} 
= \Psi \mathcal{O}^{-1} \Psi^{\dagger} \neq I_N. \label{eq:PMO_O}
\end{eqnarray} 
Here we define $P_{\rm MO}$ which is the projector to the vector space spanned by
the orthonormalized MOs $\Psi^\prime$. 
Note that the relation $P_{\rm MO}^2 = P_{\rm MO}$ holds, which is the generic requirement of the projector.

The co-space of the space spanned by the MOs corresponds to the zero-energy flat-band.
To be concrete, the projector of the zero-energy flat bands is given as
\begin{eqnarray}
P_0 = I_N - P_{\rm MO}. \label{eq:comple}
\end{eqnarray}
One can see that the followings hold:
\begin{eqnarray}
P_0 ^2 = P_0,
\end{eqnarray}
and 
\begin{eqnarray}
P_0 P_{\rm MO} &=& (I_N - P_{\rm MO}) P_{\rm MO} = P_{\rm MO} -P_{\rm MO} ^2 = 0, \nonumber  \\
P_{\rm MO} P_0  &=& P_{\rm MO}(I_N - P_{\rm MO})   = P_{\rm MO} -P_{\rm MO} ^2 = 0.
\end{eqnarray}

The rank of $P_{\rm MO}$ is $M$ and that of $P_0$ is 
$Z = N-M$ due to the assumption that det$\mathcal{O} \neq 0$.~\footnote{We emphasize that this does not mean the number of zero modes is $Z$. Indeed, additional zero modes can
appear when $\varepsilon_n = 0$ for some $n$. In that case, det$h = 0$.}.
This means the eigenspace of $P_0$ is spanned by $Z$ orthonormalized vectors $\bm{\varphi}_1, \cdots, \bm{\varphi}_Z$.
Defining 
\begin{eqnarray}
\Psi_0 := \left( \bm{\varphi}_1, \cdots, \bm{\varphi}_Z\right),
\end{eqnarray} 
by which $P_0$ can be written as
\begin{eqnarray}
P_0 = \Psi_0 \Psi_0^\dagger,
\end{eqnarray}
we have 
\begin{eqnarray}
P_0 \Psi_0 = \Psi_0, \label{eq:psi0_prop1}
\end{eqnarray}
and 
\begin{eqnarray}
\Psi^{\dagger}_0 \Psi_0 = I_Z.
\end{eqnarray}
Equation (\ref{eq:psi0_prop1}) leads to 
\begin{eqnarray}
P_{\rm MO} \Psi_0 =P_{\rm MO}   P_0 \Psi_0 =0.
\end{eqnarray}
This implies
\begin{eqnarray}
\bm{\varphi}_i^\dagger \cdot \bm{\psi}^\prime_j = 0 \label{eq:orthogonal}
\end{eqnarray}
for $i = 1,\cdots, Z$ and $j = 1, \cdots, M$;
here we have introduced $\bm{\psi}_j^\prime$ such that
\begin{eqnarray}
\Psi^{\prime} = \left( \bm{\psi}^\prime_1, \cdots, \bm{\psi}^\prime_M \right). 
\end{eqnarray}

Now, let us define a set of zero-mode fermions as~\footnote{Note that $z_i$ zero-mode fermions are not necessarily compactly supported localized states; 
as a concrete example, see Ref.~\cite{Huber2010}.  
Generally the zero mode has an extended tail. Also, in this sense, the zero-mode states are not necessarily compact localized states~\cite{Leykam2018}.}
\begin{eqnarray}
\hat{z}^\dagger_i =  \hat{\bm{c}}^\dagger \bm{\varphi}_i.
\end{eqnarray}
They satisfy the anti-commutation relation:
\begin{eqnarray}
\left\{\hat{z}_i,\hat{z}_j^\dagger \right\} = \sum_{k,l} [\bm{\varphi}^\dagger_i]_{k} [\bm{\varphi}_j]_{l} \left\{\hat{c}_k, \hat{c}_l^\dagger \right\} 
=  \bm{\varphi}^\dagger_i  \bm{\varphi}_j = \delta_{i,j}.
\end{eqnarray}
We note that $\left\{\hat{z}_i, \hat{C}^{\prime \dagger}_j \right\}= 0$ due to (\ref{eq:orthogonal});
this also implies $\left\{\hat{z}_i, \hat{C}^{\dagger}_j \right\}= 0$ because of (\ref{eq:C_Cp}).

\section{Interaction and many-body ground state \label{sec:int}}
In the following, we consider the case where $\varepsilon_n >0$ for all $n$.
In other words, the number of zero modes for $\hat{\mathcal{H}}_0$ is $Z$ and they are given as $z^\dagger_i$ $(i=1, \cdots, Z)$.
Let $\ket{t}$ be an arbitrary many-body state.
Noting 
\begin{eqnarray}
\bra{t} \hat{\mathcal{H}}_0 \ket{t} = \sum_{n=1}^{M} \varepsilon_n || \hat{C}^{\prime}_{n} \ket{t}||^2 \geq 0,
\end{eqnarray}
we see that $\hat{\mathcal{H}}_0$ is positive semi-definite.

To construct the exactly-solvable model with interactions, 
we consider the interaction of the form: 
\begin{eqnarray}
\hat{\mathcal{H}}_{\rm int} = \sum_{i,j} V_{ij} \hat{C}_i^\dagger \hat{C}_j^\dagger \hat{C}_j \hat{C}_i, \hspace{1mm} V_{ij} \geq 0, \label{eq:Hint}
\end{eqnarray}
which is the interaction among MOs. 
This interaction Hamiltonian is positive semi-definite as well,
since 
\begin{eqnarray}
\bra{t} \hat{\mathcal{H}}_{\rm int} \ket{t}  = \sum_{i,j} V_{ij} || \hat{C}_j \hat{C}_i \ket{t} ||^2 \geq 0.
\end{eqnarray}

The $Z$-particle many-body ground state 
of the Hamiltonian $\hat{\mathcal{H}} = \hat{\mathcal{H}}_0 + 
\hat{\mathcal{H}}_{\rm int}$ is given by 
\begin{eqnarray}
\ket{z} = \prod_{i=1}^{Z} \hat{z}_i^\dagger \ket{0}, \hspace{1mm} \langle z | z \rangle = 1,
\end{eqnarray}
where $\ket{0}$ is a vacuum of $\hat{c}_i$:
\begin{eqnarray}
\hat{c}_i \ket{0} = 0
\end{eqnarray}
for all $i = 1, \cdots, N$. 
The state $\ket{z}$ is the zero-energy eigenstate of $\hat{\mathcal{H}}$ because of
the anti-commutation relation between $\hat{z}$'s and ($\hat{C}$)'s.
More precisely, $\ket{z}$ satisfies $\hat{\mathcal{H}}_0 \ket{z} = 0$ and $\hat{\mathcal{H}}_{\rm int} \ket{z} = 0$ simultaneously.
Note that similar exact many-body eigenstates with vanishing interaction energy have been discussed in several contexts~\cite{Tasaki1998,Wu2007,Kuno2020_scar,Kuno2021,Hamamoto2012}.
As $\hat{\mathcal{H}}$ is positive semi-definite, $\ket{z}$ 
is a ground state for $\hat{\mathcal{H}}$. 
Also, $\ket{z}$ is the unique ground state 
for the $Z$-particle system, since $\ket{z}$ is the only state 
that satisfies $\bra{z} \hat{\mathcal{H}}_0 \ket{z} = 0$. 
It is also worth mentioning that, even when the conditions $\varepsilon_n > 0$ and $V_{i,j} \geq 0$ are relaxed, 
$\ket{z}$ is an exact eigenstate of $\hat{\mathcal{H}}_0 + \hat{\mathcal{H}}_{\rm int}$.
In fact, as we will show in Sec.~\ref{sec:ex}, $\ket{z}$ has short-range correlations and the entanglement entropy for $\ket{z}$ obeys an exact area law for a typical model,
which implies that $\ket{z}$ becomes a quantum scar when it is embedded in the middle of the many-body spectrum.

The first excited state can also be constructed in a straightforward manner.
It has (at least)~\footnote{If the bottom of the finite-energy band is degenerate, 
the degeneracy of the first excited states increases accordingly.} $Z$-fold degeneracy, labeled by $k = 1, \cdots Z$,
 and is explicitly given as
\begin{eqnarray}
\ket{\Psi^{(1;k)}} = \hat{C}_1^{\prime \dagger} \ket{(z;k)},
\end{eqnarray}
where 
\begin{eqnarray}
\ket{(z;k)} := \hat{z}^{\dagger}_1 \cdots \hat{z}^{\dagger}_{k-1} \hat{z}^\dagger_{k+1} \cdots \hat{z}^\dagger_Z \ket{0}.
\end{eqnarray}
The states $\ket{\Psi^{(1;k)}}$ are eigenstates of $\hat{\mathcal{H}}$,
since $\hat{\mathcal{H}}_0 \ket{\Psi^{(1;k)}} = \varepsilon_{1} \ket{\Psi^{(1;k)}}$
and $\hat{\mathcal{H}}_{\mathrm{int}} \ket{\Psi^{(1;k)}} = 0$,
meaning that the eigenvalue is independent of $V_{i,j}$.
Note that, for the same reason, the state $\ket{\Psi^{(n;k)}} = \hat{C}_n^{\prime \dagger} \ket{(z;k)}$ is an exact eigenstate of $\hat{\mathcal{H}}$
whose eigenenergy is $\varepsilon_{n}$.

A proof that $\ket{\Psi^{(1;k)}}$ are the first excited states is given as follows. 
To begin with, we note that $\ket{\Psi^{(1;k)}}$ are the first excited states for the non-interacting case (i.e., for $\hat{\mathcal{H}}_0$).
Let $\ket{\Xi}$ be a $Z$-particle eigenstate of $\hat{\mathcal{H}}_0 +\hat{\mathcal{H}}_{\mathrm{int}}$
that is orthogonal to $\ket{z}$ and $\ket{\Psi^{(1;k)}}$ (i.e., $\langle z \ket{\Xi}= \langle \Psi^{(1;k)} \ket{\Xi } = 0$).
Then, $\ket{\Xi}$ can be expanded by the orthonormalized $Z$-particle eigenstates of $\hat{\mathcal{H}}_0$, $\ket{\Psi_n}$, as 
$\ket{\Xi}= \sum_n \alpha_n \ket{\Psi_n}$. 
Here $\ket{\Psi_n}$ satisfies $\hat{\mathcal{H}}_0 \ket{\Psi_n} = 
E_n  \ket{\Psi_n}$ and $\langle z \ket{\Psi_n} = \langle \Psi^{(1;k)} \ket{\Psi_n } = 0$.
The coefficients $\alpha_n$ satisfy $\sum_{n} |\alpha_n|^2 = 1$. 
From the fact that $\ket{\Psi^{(1;k)}}$ are the first excited states of $\hat{\mathcal{H}}_0$, we have $E_n > \varepsilon_1$.
(We assume that the bottom of the finite-energy band is non-degenerate, but relaxing this assumption is straightforward.)
Then, we find $\bra{\Xi}\hat{\mathcal{H}}_0 \ket{\Xi}= \sum_n E_n |\alpha_n|^2  > \sum_n \varepsilon_1 |\alpha_n|^2 = \varepsilon_1$.
Further, as $\hat{\mathcal{H}}_{\mathrm{int}}$ is positive semi-definite, we have $\bra{\Xi}\hat{\mathcal{H}}_{\rm int} \ket{\Xi}\geq 0$.
These lead to $\bra{\Xi}\hat{\mathcal{H}}_0 +\hat{\mathcal{H}}_{\rm int} \ket{\Xi} >  \varepsilon_1$, which means that 
the energy of $\ket{\Xi}$ is greater than that of $\ket{\Psi^{(1;k)}}$.
Therefore, $\ket{\Psi^{(1;k)}}$ 
are the first excited states for $\hat{\mathcal{H}}_0 +\hat{\mathcal{H}}_{\rm int}$ and corresponding
energy gap is $\Delta = \varepsilon_{1}$, which does not depend on $V_{ij}$.

\section{Correlation function \label{sec:corr}}
The correlation function of the fermions at 
$i$ and $j$ with respect to the ground state is
\begin{eqnarray}
g_{ij} = \langle \hat{c}^\dagger_i \hat{c}_j \rangle = \bra{z} \hat{c}^\dagger_i \hat{c}_j \ket{z}.
\end{eqnarray}
Noting 
\begin{eqnarray}
\left\{ \hat{c}_i,\hat{z}^\dagger_k \right\} &=& [\bm{\varphi}_k]_i,
\end{eqnarray}
and thus
\begin{eqnarray}
\hat{c}_i \ket{z} = \sum_{k=1}^{Z} (-1)^{k-1} [\bm{\varphi}_k]_i \ket{(z;k)},
\end{eqnarray}
we have 
\begin{eqnarray}
g_{ij} = \sum_{k,k^\prime = 1}^{Z} (-1)^{k+k^\prime} [\bm{\varphi}^\dagger_k]_i[ \bm{\varphi}_{k^\prime}]_j \langle(z;k)|(z;k^\prime) \rangle
= \sum_{k = 1}^{Z} [\bm{\varphi}^\dagger_k]_i[ \bm{\varphi}_{k}]_j  = [\Psi_0  \Psi_0^\dagger]_{ji} = [P_0]_{ji}. \nonumber \\ \label{eq:corr_P0}
\end{eqnarray}
Using Eqs.~(\ref{eq:PMO_O}) and (\ref{eq:comple}),
we obtain
\begin{eqnarray}
g_{ij} = [I_N -P _{\rm MO}]_{ji} = \delta_{i,j} - [\Psi \mathcal{O}^{-1} \Psi^\dagger]_{ji}. \label{eq:corr}
\end{eqnarray}
Note that the expression of Eq.~(\ref{eq:corr_P0})
was obtained by Mielke in the study of flat-band ferromagnetism~\cite{Mielke1993,Mielke1999}.
In this regard, the merit of using Eq.~(\ref{eq:corr}) is that one can obtain the correlation function 
without deriving the explicit form of $P_0$, i.e., without diagonalizing the single-particle Hamiltonian $\hat{\mathcal{H}}_0$.
This is advantageous when $\hat{\mathcal{H}}_0$ is not translationally invariant due to disorders~\cite{Hatsugai2021}.
We also emphasize that the result does not depend on the interaction $V_{i,j}$,
and that the higher-order correlation function can be obtained by using the Wick's theorem. 

\section{Analytic calculation in one-dimensional systems \label{sec:ex}}
In this section, we study concrete examples.
Specifically, we focus on the one-dimensional models,
namely the saw-tooth-lattice model and the diamond-chain model, 
where the analytic form of the correlation function can be obtained. 

\subsection{Saw-tooth lattice \label{sec:sawtooth}}
%------------------------------------------------------------------%
\begin{figure}[!htb]
\begin{center}
\includegraphics[width = 0.95\linewidth]{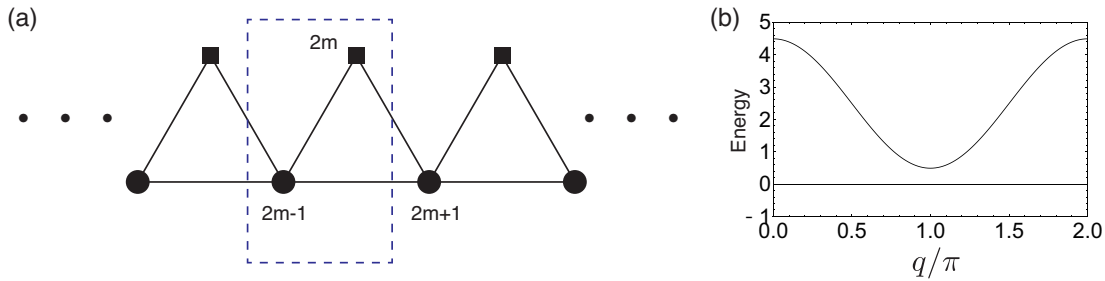}
\caption{(a) The saw-tooth lattice. (b) The band dispersion of the Hamiltonian of Eq.~(\ref{eq:Ham_ST}) for $(a_1,a_2,a_3) = (1,0.7,1)$. }
  \label{fig:ST}
 \end{center}
\end{figure}
%------------------------------------------------------------------%
We first study the saw-tooth lattice [Fig.~\ref{fig:ST}(a)], 
whose single-particle Hamiltonian is
\begin{eqnarray}
\hat{\mathcal{H}}_0 &=& \sum_{m=1}^{L} \left[a_1^\ast a_2  \hat{c}^{\dagger}_{2m-1} \hat{c}_{2m} + a_2^\ast a_3  \hat{c}^{\dagger}_{2m} \hat{c}_{2m + 1} +a_3^\ast a_1  \hat{c}^{\dagger}_{2m-1} \hat{c}_{2m+1}\right]   \nonumber \\
&+& (\mathrm{h.c.}) \nonumber \\
&+&\sum_{m=1}^{L} \left[ (|a_1|^2+|a_3|^2) \hat{c}^{\dagger}_{2m-1} \hat{c}_{2m-1} + |a_2|^2 \hat{c}^{\dagger}_{2m} \hat{c}_{2m} \right]. \label{eq:Ham_ST}
\end{eqnarray}
Here, $a_1, a_2, a_3 \in \mathbb{C}$ are the parameters.
Note that the periodic boundary condition is imposed as $\hat{c}_{2L+j}  =\hat{c}_{j}$.
Due to the translational invariance, we can perform the Fourier transformation,
and derive the band structure.
In Fig.~\ref{fig:ST}(b), we plot the band structure for a representative set of parameters. 
We see that the first band is the zero-energy flat-band, and there is a finite band gap between the first and the second bands. 

In this model, the MO is set as 
\begin{eqnarray}
\hat{C}_m = a_1 \hat{c}_{2m-1} + a_2 \hat{c}_{2m} + a_3 \hat{c}_{2m+1},
\end{eqnarray}
by which the single-particle Hamiltonian can be written as
\begin{eqnarray}
\hat{\mathcal{H}}_0 = \sum_{m=1}^{L} \hat{C}^{\dagger}_m \hat{C}_m.
\end{eqnarray}
Corresponding $\Psi^\dagger$ and $\mathcal{O}$ have the forms,
\begin{eqnarray}
[\Psi^{\dagger}]_{ij} = a_1 \delta_{j,2i-1} +a_2\delta_{j,2i} + a_3\delta_{j,2i+1},
\label{eq:51}
\end{eqnarray}
and
\begin{eqnarray}
[\mathcal{O}]_{ij} = (|a_1|^2 +|a_2|^2 + |a_3|^2) \delta_{i,j} +a_1^\ast a_3\delta_{i,j-1} + a_1a_3^\ast \delta_{i,j+1},
\label{eq:52}
\end{eqnarray}
respectively.

Here we focus on a many-body ground state at half-filling.
To obtain the correlation function of Eq.~(\ref{eq:corr}),
it is helpful to utilize the momentum-space representation.
Namely, we define a $L \times L$ unitary matrix $U_L$:
\begin{eqnarray}
[U_L]_{n, m} = \frac{1}{\sqrt{L}}e^{i q_n m},
\end{eqnarray}
and a $2L \times 2L$ unitary matrix $U_{2L}$: 
\begin{eqnarray}
[U_{2L}]_{2n-1,2m-1} = [U_{2L} ]_{2n, 2m} 
= \frac{1}{\sqrt{L}} e^{i q_n m}, 
[U_{2L}]_{2n-1,2m} = [U_{2L} ]_{2n,2m-1}=0, \nonumber \\ \label{eq:uniary}
\end{eqnarray}
for $n, m=1, \cdots,L$,  
where $q_{n} = \frac{2\pi}{L}\cdot (n-1)$.
Then, we have 
\begin{eqnarray}
U_{L} \Psi^\dagger U_{2L}^\dagger = \Psi^\dagger_{q_1} \oplus \Psi^\dagger_{q_2} \oplus \cdots \oplus \Psi^\dagger_{q_L}, \label{eq:psi_unitary}
\end{eqnarray} 
with $\Psi^\dagger_{q_n}$ being the $1 \times 2 $ matrix:
\begin{eqnarray}
\Psi^\dagger_{q_n}
= \begin{pmatrix}
a_1 + a_3 e^{- i q_n} & a_2 \\
\end{pmatrix}.
\end{eqnarray}
For later use, we define $A_{q_n} = a_1 + a_3 e^{- i q_n}$, $B_{q_n} = a_2$. 
From Eq.~(\ref{eq:psi_unitary}), we have a block-diagonal form of the overlap matrix:  
\begin{eqnarray}
U_{L} \mathcal{O} U_{L}^\dagger = \mathrm{diag}\left( \mathcal{O}_{q_1},  \cdots \mathcal{O}_{q_L} \right), \label{eq:overlap_unitary}
\end{eqnarray} 
where 
\begin{eqnarray}
\mathcal{O}_{q_n} = \Psi^\dagger_{q_n}  \Psi_{q_n} 
= |A_{q_n}|^2+  |B_{q_n}|^2. 
\end{eqnarray} 
Combining Eqs.~(\ref{eq:psi_unitary}) and (\ref{eq:overlap_unitary}),
we have
\begin{eqnarray}
\Psi \mathcal{O}^{-1} \Psi^\dagger = U_{2L}^\dagger \left[ \oplus_{n=1}^{L} P_{\mathrm{MO},q_n} \right] U_{2L}, 
\end{eqnarray}
with 
\begin{eqnarray}
P_{\mathrm{MO},q_n} = \frac{1}{|A_{q_n}|^2 + |B_{q_n}|^2 } 
\begin{pmatrix}
|A_{q_n}|^2 & A^\ast_{q_n} B_{q_n}\\
A_{q_n}  B^\ast_{q_n} &|B_{q_n}|^2 \\
\end{pmatrix}.
\end{eqnarray}

For the calculation of the correlation function $g_{i,j}$, we focus on the case of $i = 2m-1$ and $j  =2m^\prime-1$; calculations for the other cases can be carried out in the same manner. 
In this case, we have 
\begin{eqnarray}
g_{2m-1,2m^\prime-1} &=& \delta_{m,m^\prime} - \frac{1}{L} \sum_{n=1}^{L} e^{iq_n (m-m^\prime)}  \frac{|A_{q_n}|^2 }{|A_{q_n}|^2 + |B_{q_n}|^2} \nonumber \\
&=&\delta_{m,m^\prime} - \frac{1}{L} \sum_{n=1}^{L} e^{iq_n (m-m^\prime)} \left[1- \frac{|B_{q_n}|^2 }{|A_{q_n}|^2 + |B_{q_n}|^2} \right] \nonumber \\
&=& \frac{1}{L} \sum_{n=1}^{L} e^{iq_n (m-m^\prime)}  \frac{|B_{q_n}|^2 }{|A_{q_n}|^2 + |B_{q_n}|^2}.
\label{eq:corr_triangle}
\end{eqnarray}
To obtain the third line of Eq.~(\ref{eq:corr_triangle}), we have used $\sum_{n=1}^{L} e^{iq_n (m-m^\prime)}  = L \delta_{m,m^\prime}$.

%------------------------------------------------------------------%
\begin{figure}[!htb]
\begin{center}
\includegraphics[width = 0.75\linewidth]{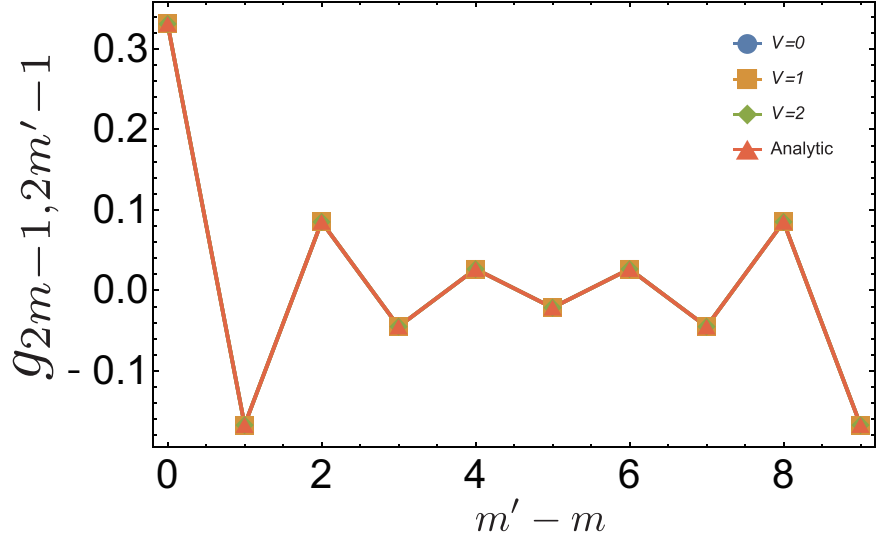}
\caption{The correlation function $g_{2m-1,2m^\prime-1}$ for $L=10$ and 10 particles.
Blue circles, orange squares, and green diamonds are, respectively, the numerical results for $V=0$, $V=1$, and $V=2$.
Red triangles are the analytic result of Eq.~(\ref{eq:corr_triangle}), which shows a good agreement with the numerical results. 
}
  \label{fig:ST_correlation}
 \end{center}
\end{figure}
%------------------------------------------------------------------%
The above expression of the correlation function is exact, 
but we perform a numerical demonstration employing the exact diagonalization to check it.
As for the interaction term, we consider $\hat{H}_{\rm int} = V \sum_m \hat{C}_m^\dagger \hat{C}_{m+1}^\dagger \hat{C}_{m+1} \hat{C}_m$.
To obtain the site-basis representation of the interaction term of 
Eq.~(\ref{eq:Hint}), we use a numerical code in Dirac-Q~\cite{Wright2013}.
For the exact diagonalization, we use the quantum lattice-model solver $\mathcal{H}\Phi$~\cite{Kawamura2017}.
In Fig.~\ref{fig:ST_correlation}, the correlation function for the system with $L=10$ (the number of sites is 20) is shown for several values of $V$.
We see that the correlation function is indeed independent of $V$.
We also see that the numerical results show a good agreement with the analytic one.
%------------------------------------------------------------------%
\begin{figure}[!htb]
\begin{center}
\includegraphics[width = 0.7\linewidth]{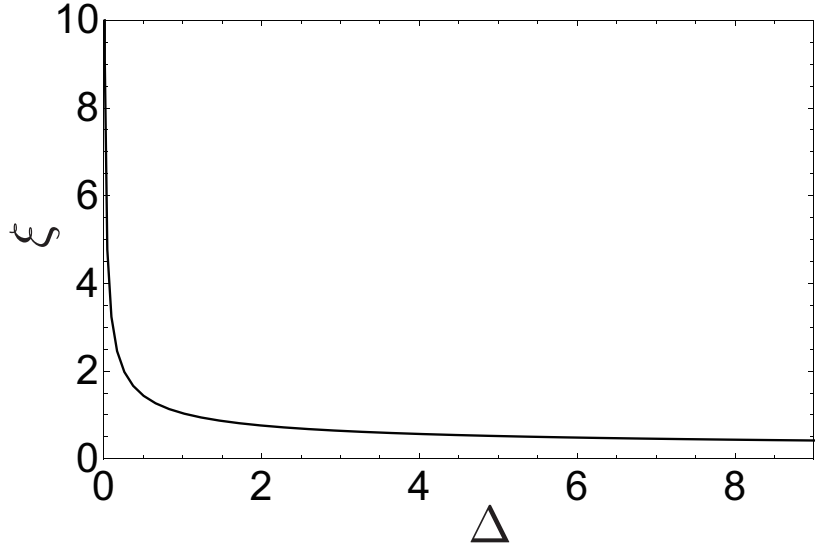}
\caption{The energy gap versus the correlation length for $a_1  =a_3 = 1$.}
  \label{fig:gap_corr}
 \end{center}
\end{figure}
%------------------------------------------------------------------%

We further evaluate the correlation length 
in the limit of $L\rightarrow \infty$. 
For $L \rightarrow \infty$, we can replace  $\frac{1}{L}  \sum_{n=1}^{L}$ with $\int_0^{2\pi} \frac{d q}{2\pi}$.
Then, we have  
\begin{eqnarray}
g_{2m-1,2m^\prime-1}  = \int_0^{2\pi} \frac{dq}{2\pi} \hspace{0.5mm}e^{i q x}  \frac{|a_2|^2}{T(1 + \epsilon e^{iq} + \epsilon^\ast e^{-iq})}, \label{eq:corr_triangle2}
\end{eqnarray}
where we have introduced $x:= m-m^\prime$,
$T:= |a_1|^2 + |a_2|^2 + |a_3|^2$, and $\epsilon:= (a_1a_3^\ast)/T$. 
Up to the coefficient $ \frac{|a_2|^2}{T}$, the right-hand side of Eq.~(\ref{eq:corr_triangle2}) can be calculated by 
replacing the variable as $z = e^{i( q+\mathrm{arg}\epsilon)}$:
\begin{eqnarray}
&\int_0^{2\pi} \frac{dq}{2\pi} \hspace{0.5mm}e^{i q x}\frac{1}{1 + \epsilon e^{iq} + \epsilon^\ast e^{-iq}} \nonumber \\
=& \frac{1}{2\pi i} \oint_{|z|=1} \frac{dz}{z} \hspace{0.5mm} \frac{z^x e^{-i (\mathrm{arg}\epsilon) x}}{1 + |\epsilon| (z + z^{-1})} \nonumber \\
=& \frac{1}{2\pi i |\epsilon|} \oint_{|z|=1}  dz  \hspace{0.5mm}  \frac{z^{x} e^{-i (\mathrm{arg}\epsilon) x} }{(z-z_<) (z-z_>)}\nonumber \\
=& \frac{(z_<)^x e^{-i (\mathrm{arg}\epsilon) x} }{|\epsilon| (z_<-z_>)} \nonumber \\
=& \frac{e^{-i (\mathrm{arg}\epsilon) x} (z_<)^x}{\sqrt{1-4|\epsilon|^2}},
\end{eqnarray}
where $z_{<, >}$ are the solutions of the quadratic equation $z^2 +|\epsilon|^{-1}z + 1=0$, i.e., 
\begin{eqnarray}
z_{>} = \frac{1}{2|\epsilon|} \left( -1- \sqrt{1- 4|\epsilon|^2} \right),
\end{eqnarray}
and 
\begin{eqnarray}
z_{<} = \frac{1}{2|\epsilon|} \left(  -1+ \sqrt{1- 4|\epsilon|^2}  \right),
\end{eqnarray}
satisfying $|z_<| < 1$ and $|z_>| > 1$. 
Therefore, the correlation function is given as
\begin{eqnarray}
g_{2m-1,2m^\prime-1} = \frac{|a_2|^2 e^{-i (\mathrm{arg}\epsilon)(m-m^\prime) }  (z_<)^{(m-m^\prime)}}{T\sqrt{1-4|\epsilon|^2}}. \label{eq:corr_sawtooth}
\end{eqnarray}
This implies the exponential decay of the correlation function. 
The correlation length $\xi$, by which the correlation function is written as 
$g_{(2m-1),(2m^\prime-1)} \sim e^{-|m-m^\prime|/\xi}$, is 
\begin{eqnarray}
\xi = -\left(\log|z_<|\right)^{-1} =  - \left(\log \frac{1 - \sqrt{1-4 |\epsilon|^2 }}{2|\epsilon|}\right)^{-1}.
\end{eqnarray}
In Fig.~\ref{fig:gap_corr}, we plot the energy gap $\Delta$ and the correlation length, fixing $a_1 = a_3 =1$ and varying $a_2$.
In this case, the energy gap is given as $\Delta = (a_2)^2$. 
We see that $\xi$ is diverging for $\Delta \rightarrow 0$ and that
$\xi$ decreases upon increasing $\Delta$.
We note that the correlation function for $\Delta = 0$ is vanishing due to the factor $|a_2|^2$ in Eq.~(\ref{eq:corr_sawtooth}).
This is accounted for as follows. For $a_2 = 0$, 
the even-numbered sites are completely decoupled from the odd-numbered sites, and the atomic states at the even-numbered sites form the flat-band. Then, the odd-numbered sites are empty for the ground state, resulting in the vanishing of the correlation function.
%------------------------------------------------------------------%
\begin{figure}[!htb]
\begin{center}
\includegraphics[width = 0.6\linewidth]{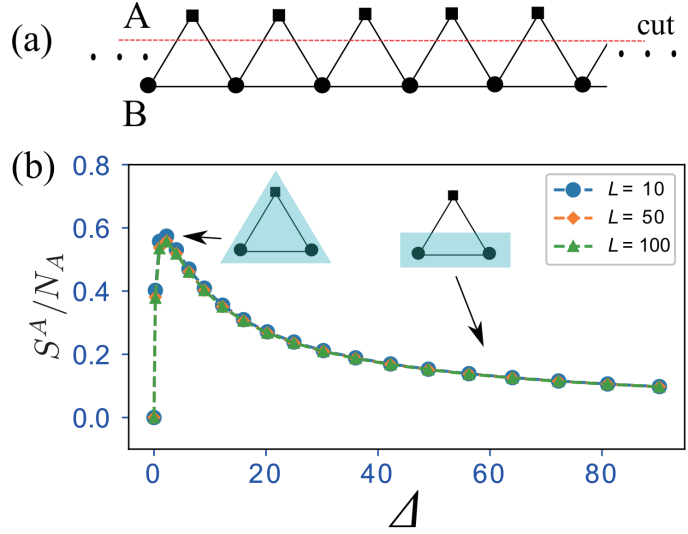}
\caption{(a) The entanglement cut for the calculation of the entanglement entropy in the saw-tooth lattice.
(b) The entanglement entropy for the A subsystem versus the energy gap $\Delta$ for $a_1  =a_3 = 1$. The inset schematic figures show schematic Wannier functions. The blue shaded regime indicates the schematic finite amplitude of the Wannier function. The periodic boundary condition is used.}
  \label{fig:ES}
 \end{center}
\end{figure}
%------------------------------------------------------------------%

Furthermore, we show another useful application of the correlation function analytically obtained by Eq.~(\ref{eq:corr}). That is, the entanglement entropy can be obtained directly from the correlation function.
The reduced density matrix for a certain subsystem (A-subsystem) denoted by $\rho^A$ can be extracted from the correlation function \cite{Peschel2003,Ryu2006,Peschel2009}
\begin{eqnarray}
\rho^A_{i,j\in A}=g_{i,j \in A}.
\end{eqnarray}
Here, the eigenvalues of $\rho^A$ are the entanglement spectra denoted by $\lambda_{n}$ ($n=1,2,\cdots, N_A$, where $N_A$ is the number of lattice sites of the A-subsystem). 
Then, from the values of $\lambda_{n}$, the entanglement entropy is given by \cite{Ryu2006},
\begin{eqnarray}
S^A=-\sum^{N_A}_{n=1}[\lambda_{n}\ln \lambda_{n}+(1-\lambda_{n})\ln (1-\lambda_{n})]. \label{eq:EE_SA}
\end{eqnarray}

We can extract the entanglement property of the many-body ground state at half-filling in the saw-tooth lattice.
By substituting Eq.~(\ref{eq:51}) and Eq.~(\ref{eq:52}) into Eq.~(\ref{eq:corr}) and setting the entanglement cut 
where the A-subsystem includes all even-numbered sites as shown in Fig.~\ref{fig:ES}, we calculated $S^{A}$. 
Here the eigenvalues of the correlation matrix, $\lambda_n$, are computed numerically.
We note, however, that this bipartition is translationally symmetric, and thus $S^{A}$ can be computed by using the momentum space representation;
see Appendix~\ref{app:ee} for details.
The result of $S^A$ versus the energy gap $\Delta(=|a_2|^2)$ for $a_1=a_3=1$ is shown in Fig.~\ref{fig:ES}(b).
The entanglement entropy between even ($2m$) sites and odd ($2m+1$) sites depends on the size of the gap $\Delta$. 
The entanglement entropy increases for small $\Delta$ to a peak at $S^A/N_A \sim \ln 2$, and then decays as increasing $\Delta$.
Actually, this behavior is related to the form of the Wannier function of the flat band. From the Hamiltonian $H_0$ of Eq.~(\ref{eq:Ham_ST}), the Wannier function can be calculated. 
At the saturation point of $S^{A}$, $\Delta\sim 1.8$, the Wannier function is mainly localized on
three sites as shown in the inset of Fig.~\ref{fig:ES}(b). Here, the two odd-numbered sites and the single even-numbered site are highly entangled. 
On the other hand, for larger $\Delta$, the Wannier function is mainly localized on the two adjacent odd sites as shown in the inset of Fig.~\ref{fig:ES}(b). This means that the odd-numbered sites and even-numbered sites are weakly entangled. 
And also, the result in Fig.~\ref{fig:ES}(b) shows almost no system-size dependence.
This indicates that the entanglement of the many-body ground state shows the area-law, 
which is characteristic in generic unique gapped ground states.

\subsection{Diamond chain}
%------------------------------------------------------------------%
\begin{figure}[!htb]
\begin{center}
\includegraphics[width = 0.95\linewidth]{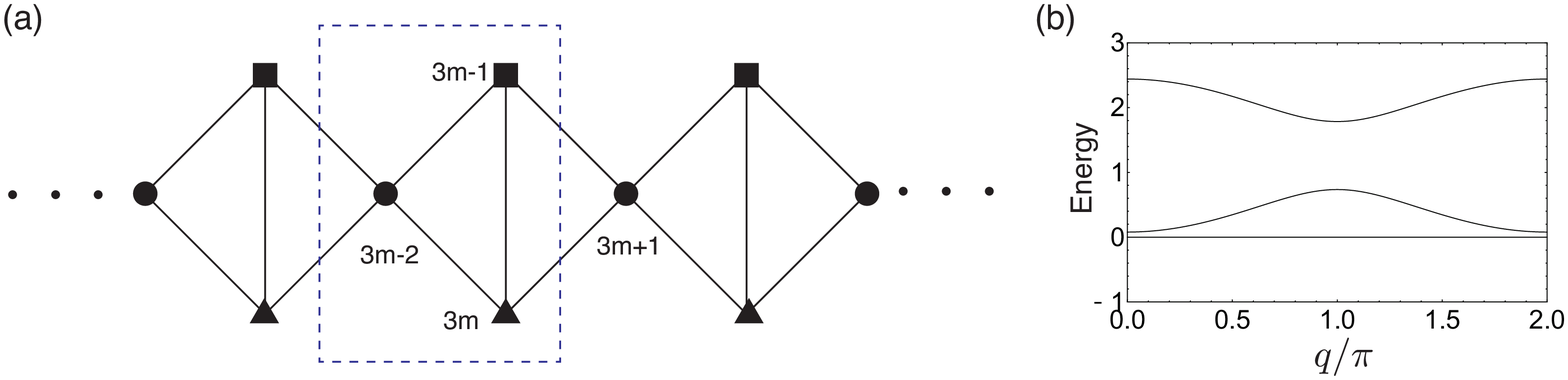}
\caption{(a) The diamond chain. (b) The band dispersion of the Hamiltonian of Eq.~(\ref{eq:Ham_Dia}) for $(a_1,a_2,a_3,a_4,a_5,a_6) = (0.5,0.8,0.8,0.5,0.5,0.7)$. }
  \label{fig:D}
 \end{center}
\end{figure}
%------------------------------------------------------------------%
The second example is the diamond chain [Fig.~\ref{fig:D}(a)].
The single-particle Hamiltonian reads
\begin{eqnarray}
\hat{\mathcal{H}}_0 &=& \sum_{m=1}^{L} [a_1^\ast a_2  \hat{c}^{\dagger}_{3m-2} \hat{c}_{3m-1} + a_1^\ast a_3 a \hat{c}^{\dagger}_{3m-2} \hat{c}_{3m} \nonumber \\
&+& (a_2^\ast a_3  + a_4^\ast a_5) \hat{c}^{\dagger}_{3m-1} \hat{c}_{3m} 
+ a_4^\ast a_6 \hat{c}^\dagger_{3m-1} \hat{c}_{3m+1} + a_5^\ast a_6 \hat{c}^\dagger_{3m} \hat{c}_{3m+1} )]\nonumber \\
&+&  (\mathrm{h.c.})\nonumber \\
&+& \sum_{m=1}^{L}\left[ \left( |a_1|^2 + |a_6|^2 \right) \hat{c}^{\dagger}_{3m-2} \hat{c}_{3m-2} 
+  \left(|a_2|^2 + |a_4|^2  \right) \hat{c}^{\dagger}_{3m-1} \hat{c}_{3m-1} 
+\left(|a_3|^2 + |a_5|^2 \right)\hat{c}^\dagger_{3m}\hat{c}_{3m}\right],  \nonumber \\ \label{eq:Ham_Dia}
\end{eqnarray}
where $a_1$-$a_6 \in \mathbb{C}$ are the parameters 
and the periodic boundary condition is imposed as $\hat{c}_{3L+j} = \hat{c}_j$. 
The band structure for the representative parameters is shown in Fig.~\ref{fig:D}(b),
where we see the zero-energy flat-band with a finite energy gap to the other bands. 

For this model, we define two species of MOs:
\begin{subequations}
\begin{eqnarray}
\hat{C}_{2m-1} = a_1 \hat{c}_{3m-2}  + a_2 \hat{c}_{3m-1} + a_3 \hat{c}_{3m}, 
\end{eqnarray}
\begin{eqnarray}
\hat{C}_{2m} = a_4 \hat{c}_{3m-1}  + a_5 \hat{c}_{3m} + a_6 \hat{c}_{3m+1}, 
\end{eqnarray}
\end{subequations}
and then the Hamiltonian can be written as
\begin{eqnarray}
\hat{\mathcal{H}}_0 &=& \sum_{m=1}^{L} \hat{C}^{\dagger}_{2m-1} \hat{C}_{2m-1} + \hat{C}^\dagger_{2m} \hat{C}_{2m}. 
\end{eqnarray}

The momentum-space representation of $\Psi^{\dagger}$ and $\mathcal{O}$ can be obtained by the same procedure as the previous subsection,
and we have
\begin{eqnarray}
\Psi^{\dagger}_{q_n} = 
\begin{pmatrix}
a_1 & a_2 & a_3 \\
a_6 e^{-iq_n} & a_4 & a_5\\
\end{pmatrix}
\end{eqnarray}
and
\begin{eqnarray}
\mathcal{O}_{q_n} =
\begin{pmatrix}
X_1  & Y_{q_n} \\
Y^\ast_{q_n} & X_2 \\
\end{pmatrix},
\end{eqnarray}
where $X_1  = |a_1|^2 + |a_2|^2 + |a_3|^2 $, $X_2 = |a_4|^2 + |a_5|^2 + |a_6|^2 $,
and $Y_{q_n} = a_1 a_6^\ast e^{iq_n}  + a_2 a_4^\ast + a_3a_5^\ast$. 
Then we find
\begin{eqnarray}
[\Psi \mathcal{O}^{-1} \Psi^\dagger]_{(3m^\prime-2)(3m-2)}
= \sum_{n=1}^{L} e^{iq_n (m-m^\prime)} 
\frac{\alpha + \beta e^{iq_n} + \beta^\ast e^{-iq_n}}{\alpha^\prime + \beta e^{iq_n} + \beta^{\ast} e^{-iq_n}},
\end{eqnarray}
where
\begin{subequations}
\begin{eqnarray}
\alpha =  |a_6|^2 X_1 + |a_1|^2 X_2 -2 |a_1|^2 |a_6|^2,
\end{eqnarray}
\begin{eqnarray}
\beta =- a_1 a_6^\ast (a_2^\ast a_4 + a_3^\ast a_5),
\end{eqnarray}
and 
\begin{eqnarray}
\alpha^\prime = 
|a_1|^2(|a_4|^2 + |a_5|^2) +|a_2|^2(|a_5|^2 + |a_6|^2) + |a_3|^2(|a_4|^2 + |a_6|^2)-2 \mathrm{Re}[ a_2a_3^\ast a_4^\ast a_5].\nonumber \\
\end{eqnarray}
\end{subequations}

We focus on a many-body ground state at 1/3-filling.
Then the remaining calculation of the correlation function follows that of the previous subsection, and we obtain the correlation function of the limit of $L\rightarrow \infty$ as
\begin{eqnarray}
g_{(3m-2)(3m^\prime-2)} = \frac{(\alpha^\prime-\alpha) \tilde{z}_<^{m-m^\prime} e^{-i (\mathrm{arg}\gamma) (m-m^\prime)}}{\alpha^\prime \sqrt{1-4|\gamma|^2}}, \nonumber \\
\end{eqnarray}
where $\gamma = \beta /\alpha^\prime$ and 
\begin{eqnarray}
\tilde{z}_< = \frac{1}{2|\gamma|} \left(-1 + \sqrt{1-4|\gamma|^2} \right).
\end{eqnarray}
We thus obtain the correlation length,
\begin{eqnarray}
\xi = - \left( \log \frac{1- \sqrt{1-4|\gamma|^2}}{2|\gamma|} \right)^{-1}.
\end{eqnarray}

\section{Summary \label{sec:summary}}
We have formulated how to calculate the correlation function in the flat-band models,
whose single-particle Hamiltonian and the interaction term are both
constructed from the MOs.
Our main result is represented by Eq.~(\ref{eq:corr}), 
which implies the correlation function depends only on the structure of 
MOs, irrespective of the details of $V_{i,j}$
(as far as the interaction term is positive semi-definite). 
This formulation of calculating the correlation function has some useful applications such as obtaining 
the correlation length and the entanglement entropy of the many-body ground state.

As concrete examples, we study the saw-tooth lattice and the diamond chain, where we presented the analytic forms of the correlation functions and the correlation length. 
In these examples, the MOs are composed of the atomic sites in a range of neighboring unit cells. 
Thus, the analytic calculation of the correlation function can be performed by using the changing the variable as $z \propto e^{iq}$ and using the residue theorem. 
It is noteworthy that we can perform the analytic calculation in the same way for the models with the MOs are composed of the atomic sites with farther unit cells.
In such cases, the correlation length is determined by 
the pole of the integrand with the absolute value being the closest to 1.

\section*{Acknowledgment}
We wish to thank H. Katsura for fruitful comments.
This work is partly supported by JSPS KAKENHI Grant 
No.~JP17H06138, 
No.~JP20K14371 (T.M.), and No.~JP21K13849 (Y.K.).

\appendix
\section{Derivation of the entanglement entropy using momentum-space representation \label{app:ee}}
In this appendix, we derive the analytic form of the entanglement entropy 
for the saw-tooth lattice model shown in Fig.~\ref{fig:ES}, derived by using the momentum-space representation.
As we have seen, the correlation function does not depend on the interaction, 
so it is sufficient to consider the single-particle Hamiltonian. 

To begin with, we introduce the Bloch Hamiltonian by performing the unitary transformation of Eq.~(\ref{eq:uniary}) to $H_0$. 
By doing so, we have
\begin{eqnarray}
U_{2L}H_0U^\dagger_{2L} = H_{0,q_1} \oplus H_{0,q_2}\oplus \cdots \oplus H_{0,q_L}
\end{eqnarray}
with 
\begin{eqnarray}
H_{0,q_n} = 
\begin{pmatrix}
|A_{q_n}|^2 &  A^\ast_{q_n} B_{q_n}\\
 A_{q_n} B^\ast_{q_n} & |B_{q_n}|^2\\  
\end{pmatrix},
\end{eqnarray}
where $A_{q_n}$ and $B_{q_n}$ are those defined in
Sec.~\ref{sec:sawtooth}.
The flat-band eigenstate is obtained 
as the zero-energy eigenvector 
of $H_{0,q_n}$, $\bm{\varphi}_{q_n}$, 
whose explicit form is 
\begin{eqnarray}
\bm{\varphi}_{q_n} = 
\frac{1}{\sqrt{|A_{q_n}|^2 + |B_{q_n}|^2}}
\begin{pmatrix}
- B_{q_n} \\
A_{q_n} \\
\end{pmatrix}.
\end{eqnarray}
Using 
\begin{eqnarray}
c_{q_{n},\mathrm{o}} = \sum_{m=1}^{L} c_{2m-1} 
e^{i q_n m}, \hspace{1mm}
c_{q_{n},\mathrm{e}} = \sum_{m=1}^{L} c_{2m} e^{i q_n m},
\end{eqnarray}
and 
\begin{eqnarray}
\bm{c}_{q_n} = 
\begin{pmatrix}
c_{q_{n},\mathrm{o}} \\c_{q_{n},\mathrm{e}} \\
\end{pmatrix},
\end{eqnarray}
where the subscripts o and e stand for the odd-numbered and even-numbered sites, respectively,
we can write the creation operator of the flat band state, $z^\dagger_{q_n}$, 
as
\begin{eqnarray}
z^\dagger_{q_n} = \bm{c}^\dagger_{q_{n}} \bm{\varphi}_{q_n}.
\end{eqnarray}
Then, the many-body ground state is written as
\begin{eqnarray}
\ket{\Psi_0} = \prod_{n=1}^{L} z^\dagger_{q_n} \ket{0}.
\end{eqnarray}

Now, the correlation function for 
the even-numbered sites are
\begin{eqnarray}
g_{2m,2m^\prime} &=& \bra{\Psi_0} c^{\dagger}_{2m} c_{2m^\prime} \ket{\Psi_0} \nonumber \\
&=& \sum_{n,n^\prime}e^{i(q_n m-q_{n^\prime} m^\prime)}
\bra{\Psi_0} c^\dagger_{q_n,\mathrm{e}}  c_{q_{n^\prime},\mathrm{e}}  \ket{\Psi_0}.
\end{eqnarray}
Noting 
\begin{eqnarray}
c_{q_{n},\mathrm{e}}  \ket{\Psi_0} = c_{q_{n},\mathrm{e}} \prod_{\ell}z^\dagger_{q_{\ell}} \ket{0}
= (-1)^{n-1} [\bm{\varphi}_{q_n}]_2 \prod_{\ell \neq n}z^\dagger_{q_{\ell}} \ket{0},
\end{eqnarray}
we have 
\begin{equation}
g_{2m,2m^\prime}  = \sum_{n}e^{i q_n(m-m^\prime)} |[\bm{\varphi}_{q_n}]_2|^2
 = \sum_{n} e^{i q_n(m-m^\prime)} 
 \frac{|A_{q_n}|^2}{|A_{q_n}|^2 + |B_{q_n}|^2 }.
\end{equation}

Having this at hand, we can compute the entanglement entropy 
$S^A$ by using the eigenvalues of the $L \times L$ matrix $g_{2m,2m^\prime}$ $(m,m^\prime = 1, \cdots, L)$.
After some algebras, 
we find that the eigenvalues 
are given as 
\begin{equation}
\lambda_n =\frac{|A_{q_n}|^2}{|A_{q_n}|^2 + |B_{q_n}|^2 }, \label{eq:lambda_ana}
\end{equation}
thus the entanglement entropy is 
by substituting $\lambda_n$ of Eq.~(\ref{eq:lambda_ana}) 
into Eq.~(\ref{eq:EE_SA}).
%------------------------------------------------------------------%
\begin{figure}[!htb]
\begin{center}
\includegraphics[width = 0.6\linewidth]{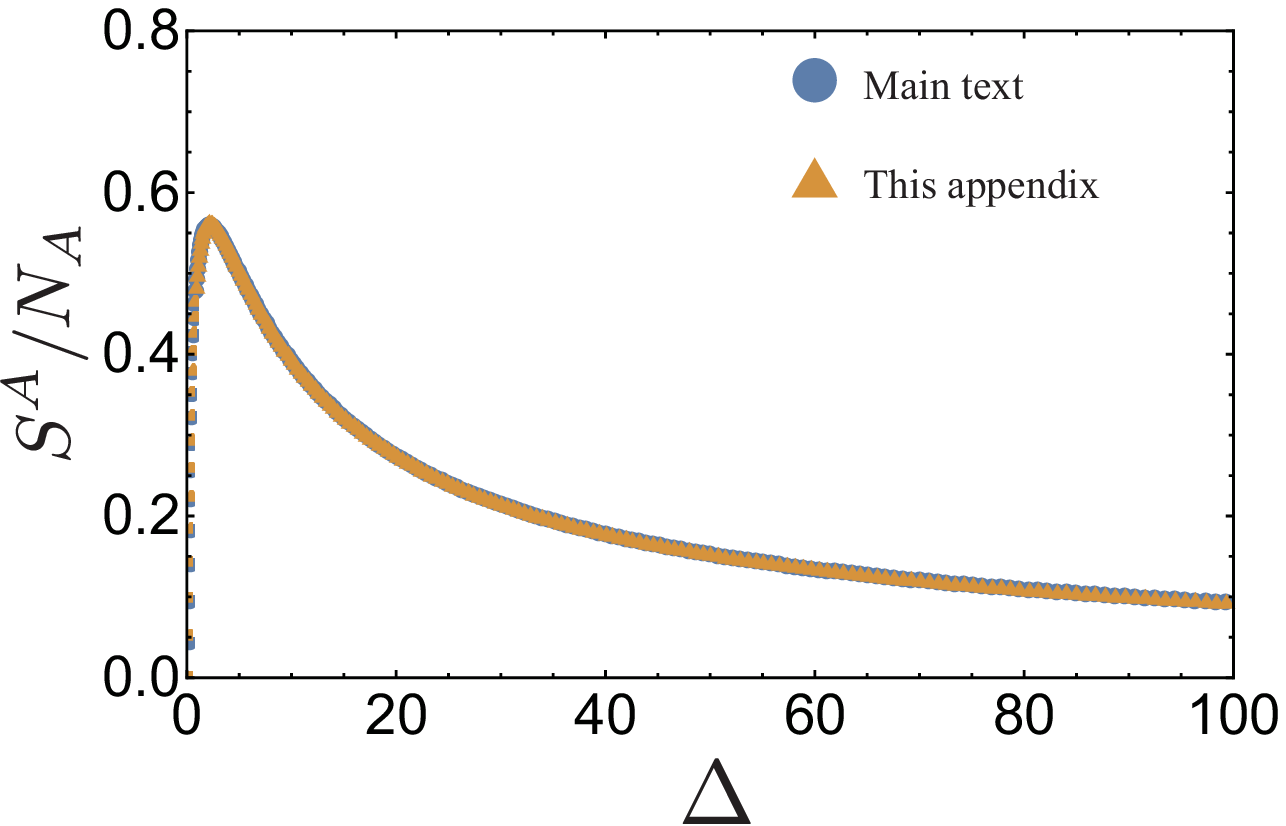}
\caption{$\Delta$ dependence of $S^A/N_A$ for $L=64$.
The blue circles and orange triangles denote the result obtained by the method described in the main text and that of analytical expression derived in the appendix, respectively.}
  \label{fig:EEana}
 \end{center}
\end{figure}
%------------------------------------------------------------------%

We now focus on the case of $a_1 = a_3 = 1$.
Figure~\ref{fig:EEana} shows 
the $\Delta$ ($=a_2^2$) dependence of $S^A/N_A$
for $L=64$, comparing the result obtained in the main text with that of the analytical expression derived in this appendix.
Clearly, they show a good agreement within the numerical accuracy.

\end{document}